\documentclass[10pt, reprint, aps, amsmath,amssymb,twocolumn]{revtex4-1}

\usepackage{epigraph}
\usepackage{graphicx}
\usepackage{dcolumn}
\usepackage{bm}
\usepackage[font=scriptsize]{caption}
\usepackage{booktabs}
\usepackage{subcaption,graphicx}

\begin{document}

\title{Towards a Simple, and Yet Accurate, Transistor Equivalent Circuit  and \\ Its Application 
to  the Analysis and Design of Discrete and Integrated Electronic Circuits}

\author{Luciano da F. Costa}
\email{ldfcosta@gmail.com, luciano@if.sc.usp.br}
\affiliation{S\~ao Carlos Institute of Physics, IFSC-USP,  S\~ao~Carlos, SP,~Brazil}

\date{05 February 2018}

\begin{abstract}
Transistors are the cornerstone of modern electronics.  Yet, their relatively complex characteristics, allied with often 
observed great parameter variation, remain a challenge for discrete and integrated electronics.  Much of transistor 
research and applications have relied on transistor models, as well as respective equivalent circuits, to be employed for 
circuit analysis and simulations.  Here,  a simple and yet accurate transistor equivalent circuit is derived, based on the 
Early effect, which involves only the voltage $V_a$ and a companion parameter $s$.  Equations are obtained for
currents and voltages in a common-emitter circuit, allowing the derivation of respective
gain functions.  These functions are found to exhibit interesting mathematical structure, with gain values 
varying almost linearly with the base current, allowing the gains to be well characterized
in terms of their average and variation values.  These results are applied to deriving a prototypic Early space 
summarizing the characteristics of transistors, enriched with recently experimentally obtained prototypes of NPN and PNP
silicon BJTs and alloy germanium transistors.  Though a trade-off between gain and
linearity is revealed, a band characterized by small values of $V_a$ stands out when aiming
at both high gain and low distortion.  The Early equivalent model was used also for studying the stability of circuits 
under voltage supply oscillations, as well as parallel combinations of transistors.  In the former
case, it was verified that more traditional approaches assuming constant current gain can yield stability factors that 
deviate substantially from those derived for the more accurate Early approach.  The equivalent circuit obtained for 
parallel combinations of transistors was shown also to closely follow the Early formulation.
\end{abstract}

\maketitle

\setlength{\epigraphwidth}{.49\textwidth}
\epigraph{\emph{``Where there is matter, there is geometry.''}}{J. Kepler}

\section{Introduction}

Developed in 1947~\cite{riordan:1997}, transistors quickly became the cornerstone of modern electronics.  Because 
of their central importance, be it in integrated or discrete circuits, these devices have become the subject of intense and 
continued research from both theoretical (e.g.~semiconductor  physics) and practical (e.g.~electronic engineering) 
points of view.  There are two main applications of transistors: digital and analog 
(e.g.~\cite{jaeger:1997,sedra:1998,boylestad:2008,horowitz:2015,streetman:2016}).  
In the former case, transistors are used as switches transitioning between two
logical levels, which underlies the area of \emph{digital electronics}.   In the latter case, transistors are typically used as linear
amplifiers, characterizing the area known as \emph{analog} (or \emph{linear}) \emph{electronics} 
(e.g.~\cite{pettit:1961,hyat:1962,analogdesign:1988,analogdesign:1991,casier:2011,carusone:2012,sedra:1998}).  
Because the many signals in nature exhibit continuous values, being therefore called \emph{analog}, linear electronics continues 
to be essential for myriad electronic applications.  Here, the challenge is to achieve efficient circuits capable of amplifying signals with the
lowest level of distortion.  This constitutes a challenging endeavor because transistors are inherently non-linear devices.  So, at least 
three alternative approaches have been typically considered in order to try obtaining linear amplifiers by using transistors: (i) to improve the
of the devices; (ii) to find the most linear operation region; and (iii) to develop circuits capable of enhancing linearity.  With the introduction 
of negative feedback in electronics by H. S. Black in 1927~\cite{black:1934}, 
alternative (iii) became the most standard and commonly applied approach for achieving practical amplifying circuits with improved
linearity.   

All the above mentioned three approaches to enhance linearity when using transistors share a common, critical aspect: they all rely critically 
on the availability of effective transistor models, representations and equivalent circuits.  As a consequence, several modeling 
and equivalent circuits have been proposed and used in linear electronics 
(e.g.~\cite{pettit:1961,hyat:1962,analogdesign:1988,analogdesign:1991,jaeger:1997,casier:2011,carusone:2012,streetman:2016}).  
An alternative transistor modeling approach was reported recently~\cite{costaearly:2017,costaearly:2018} that relies on 
the Early effect, discovered by J. M. Early in 
1952~\cite{early:1952,ziel:1968,streetman:2016}.     This effect is characterized
by the variation of the charge carrier portion of the base with the base-collector voltage.  An immediate consequence of this phenomenon
in junction transistors is that the characteristic isolines of transistors (indexed by the base current) will not cross the collector voltage 
axis at $V_C = 0$, but at a further away negative value $V_C = V_a$, which corresponds to the \emph{Early voltage}.  Though these 
facts have been known for a long time, they have not often considered for modeling transistors, except for a few works relating the 
Early voltage with the transistor output resistance $R_o$, especially in the case of FET devices (e.g.~\cite{AD:2017}).  

The methodology described in ~\cite{costaearly:2017,costaearly:2018} uses the Early effect to derive a complete transistor model 
characterized by two parameters:
the Early voltage $V_a$ and a proportionality parameter $s$ relating the characteristic isoline angles $\theta$ with the modulating
base  current $I_B$.  Perhaps the key 
element in this alternative type of transistor modeling consists in the fact that it was experimentally found for hundreds of small signal
BJTs~\cite{costaearly:2017,costaearly:2018,germanium:2018} that $\theta = s I_B$, i.e.~the isoline angles are directly proportional to 
the base current.   This linear relationship not only simplifies
the Early modeling approach, but also leads to the specially important property that both the parameters involved in the Early model 
result \emph{completely independent of the transistor/circuit operation} in the space $(V_C,I_C)$.  This contrasts sharply with the fact that
in the more traditionally applied models, the two involved parameters current gain $\beta$ and output resistance $R_o$ vary with the normal operation
of the circuit.   This has implied that, given a specific transistor, it is impossible to characterize it by a single parameter setting $(\beta, 
R_o)$, being necessary to consider maximum or average values of these parameters along the operation space, which is not accurate
because the parameters of the transistor characteristic isolines tend to vary extensively even in normal operation.  It should be recalled that
several other transistor models have been employed (e.g.~\cite{gray:1969,sedra:1998,boylestad:2008,streetman:2016}), 
some of which aimed at providing more complete representations by incorporating a larger number of components.

In spite of its recent introduction, the Early modeling approach~\cite{costaearly:2017,costaearly:2018} has already been successfully 
applied to several issues in electronics, 
including the characterization of real-world NPN and PNP silicon~\cite{costaearly:2017} and germanium~\cite{germanium:2018} 
junction devices, allowing the derivation of
a prototypical atlas of device characteristics in which NPN exhibit lower parameter variations than the PNP counterparts.  In addition,
it was found that both the NPN and PNP transistors tended to have the same average $\beta$ while differing markedly in $V_a$ and 
$s$ values, parameters that had not been usually considered.   It has also been
verified that the total harmonic distortion of transistors depends only on the parameter $s$, being independent of the Early voltage $V_a$.
Actually, the intrinsic adherence of the Early model to the geometrical structure of transistors, allied with the simplicity and accuracy of the 
proposed basic mathematical representation, paves the way to several advances in understanding, designing, and applying transistors
in integrated (e.g.~\cite{gray:1990,carusone:2012}) and discrete circuits.   
The geometry of device operation constitutes a particularly promising perspective because not only it
promotes better and more intuitive understanding of the potential behavior of a device (often in a very complete way), but it also provides
the scaffolding for deriving more effective and intuitive mathematical representations, models, and respective equivalent circuits.  
Indeed, graphical approaches have been extensively used since the beginnings of electronics, and many of the textbooks from 50's to 70's 
(e.g.~\cite{shea:1955,zimmermann:1959,pettit:1961,fontaine:1963,alley:1966,gronner:1970,tinnell:1972}) 
are characterized by extensive and systematic use of graphical presentations and developments.  Perhaps the progressive shift to 
numeric-computational modeling of device operation along the 70's and 80's has shifted a little bit this paradigm, but graphical approaches 
remain, nevertheless, an interesting resource to be considered, exhibiting potential for enhancing the available simulation resources.
So, its is not that graphic representations of device operation should be replaced by numeric-computational simulation approaches, but 
that it should be incorporated as a valuable and useful concept that could contribute to advances in simulation approaches. 
 
By having access to effective transistor models, devices operating in the so-called ``linear'' regime and having forward-biased base-emitter junction 
and inversely-biased base-collector junctions can be conveniently substituted by their respective equivalent circuits, providing a
better understanding of the circuit characteristics and promoting possibilities for further enhancements. The derivation of transistor 
equivalent circuits can benefit greatly from the Early model~\cite{costaearly:2017,costaearly:2018} intrinsic simple geometry fully compatible
with transistor operation, characterized by radiating isolines.  The main objective of the
current work is to develop such an equivalent circuit for the Early model of transistors, and illustrate its theoretical and practical 
application potentials with respect to the characterization
of important electronic properties of transistor-based circuits, including current, voltage and power gains as well
as distortion.  In addition, the proposed equivalent model is also applied to the analyses stability of circuits in presence of power
supply oscillations and parallel combinations of BJTs.

This article starts by revising a more traditional transistor modeling approaches based on the current gain $\beta$ and output resistance
$R_o$, and proceeds by presenting the Early modeling approach and deriving respective equivalent circuits, from which current and
voltage equations describing the behavior of a simplified common-emitter circuit configuration are derived.  These equations are then
employed to quantify respective current and voltage gains, exhibiting an interesting mathematical structure that implies that, at least
for the considered circuit and parameter configurations, the gains vary in very nearly to linear fashion with the modulating current base
$I_B$.  This fact allows the gains to be effectively quantified in terms of their respective average and variation values.  These results
allowed the derivation of a prototypic Early space characterizing a trade-off between gain and linearity, as well as incorporating prototypic
groups of NPN and PNP silicon and germanium devices.  The potential of the Early equivalent circuit is further illustrated with respect
to applications to the study of stability induced by voltage supply oscillations as well as for the characterization of parallel combinatios
of transistors.

\section{Real-World Transistor Characteristic Surfaces}

\begin{figure}[h!]
\centering{
\includegraphics[width=9cm]{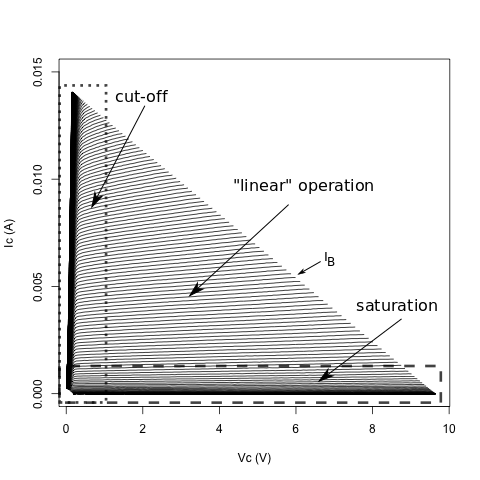}
\caption{The characteristic surface, represented by base current-indexed isolines, of a real-world PNP small signal transistor.  
                 The cut-off and saturation regions are excluded from the ``linear'' region of operation.  Observe the progressively
                  increasing slopes of the isolines with $I_B$.}
\label{fig:raw}}
\end{figure}

Real-world transistors are characterized by several specific features regarding their electronic behavior, and these features tend to
vary from one device to another.  Figure~\ref{fig:raw} illustrates the characteristic surface, represented in terms of its constituting isolines,
each one indexed by a respective base current $I_B$, of a real-world PNP small signal transistor.   This characteristic surface was
obtained experimentally by using a microprocessed acquisition system~\cite{costaearly:2017,costaearly:2018}.  The operation space corresponds
to the Cartesian coordinate system $V_C \times I_C$, where $V_C$ is the voltage current measured with reference to the ground
and $I_C$ is the collector current (positive sign corresponding to current entering the device).   The characteristic surface in this
figure was obtained for the operation region defined by $V_{C,min} = 0V$,  $V_{C,max} = 10V$,  $I_{C,min} = 0 mA$ and  $V_{C,max} = 15mA$.
The transistor was set in a simplified common emitter configuration~\cite{alley:1966,gray:1969,thomson:1976,horowitz:2015}, 
with inversely biased base-collector junction and forward-biased 
base-emitter junction, with a resistive load $R_L = 670 \Omega$ attached between 
the collector and the external voltage source $V_{CC} = 12V$.  \emph{For simplicity's sake, all
current and voltage values of PNP transistors are represented by inverse values, so as to keep the operation space in the first
quadrant, therefore achieving a unified approach with respect to NPN devices.}  

Interestingly, characteristic surfaces such as that shown in Figure~\ref{fig:raw} incorporate all information necessary to characterize
the operation of the respective device, except in cases where it has reactive components.  The most
important region of the characteristic surface is that where the $I_B$ indexed isolines tend to be straight and reasonably spaced one
another.   This region is often called the \emph{``linear'' operation region}.  The narrow vertical region next to the $I_C$ axis corresponds
to the transistor \emph{saturation} and are normally avoided during linear transistor operation.  Similarly, the narrow horizontal strip next to the
$V_C$ axis, often called the transistor \emph{cut-off} region, is characterized by coalescence of isolines and is similarly not taken into
account for linear transistor operation.  

The isolines in the  linear region are mostly straight lines with slopes that increase with $I_B$.  Recall that the slopes of these isolines 
correspond to  $1/R_o$, where $R_o$ is the output resistance of the collector.  Observe also that the isolines bend at the cut-off region, 
but are very nearly straight otherwise.  NPN transistors have similar characteristic surfaces, except for the fact that the isolines tend to 
have smaller slope~\cite{costaearly:2018}.  Because the characteristic surfaces of real-world transistors are somewhat complicated, simplified
versions are typically adopted for obtaining mathematical characterization of the transistor behavior, as well as respective equivalent
circuits.  A traditional approach to modeling the characteristic surface, as well as the respectively derived equivalent circuit, is 
revised in the next section.

\section{A Traditional Approach to Transistor Modeling} \label{sec:trad}

\begin{figure}[h!]
\centering{
\includegraphics[width=6cm]{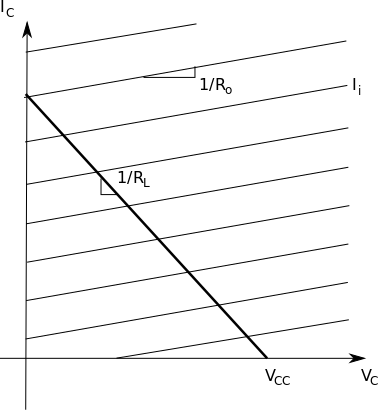}
\caption{The geometrical set-up defined by characteristic surface of a simplified NPN transistor in which all equispace isolines
defined by $I_B$ have the same slope $1/R_o$ being, therefore, parallel one another.  The load line defined by having
a load resistance $_L$ in series with a voltage source $V_{CC}$ are also included.  The state value $(V_C,I_V)$ of the 
transistor is restricted to excursioning along the load line.}
\label{fig:parallels}}
\end{figure}

Figure~\ref{fig:parallels} shows a simple NPN transistor characteristic surface in the $(V_C,I_C)$ circuit operation space, in which all 
the equispaced $I_B$-indexed isolines are assumed to have the same slope $1/R_o$, therefore being parallel one another.   This configuration
is normally used assuming that the base-collector junction is reversely biased while the base-emitter junction is forward biased, hence
the input port can be approximated by a simplified diode model as in this figure, where $R_i$ stands for its internal resistance (observe
that the ideal diode in this figure can be omitted under the above hypotheses).

Though it is know that, for most real-world transistors, the slopes actually increase with $I_B$, this fact is not considered in this model 
for simplicity's sake.   As a matter of fact, an even simpler transistor characteristic
is sometimes used in which all equispace isolines have null inclination, implying $Ro = \infty$.  Despite these simplifications, such approaches
remain interesting, as they provide insights about how more idealized transistors would behave, and can also be used in introductory
didactic syllabuses.

In this section, we review the analysis of the electronic properties of the ideal transistor represented by the characteristic
surface in Figure~\ref{fig:parallels}.  First, we obtain the current and voltage equations for the input and output ports, and then use these
equations in order to derive the current, voltage and power gain, and discuss linearity.

\begin{figure}[h!]
\centering{
\includegraphics[width=8cm]{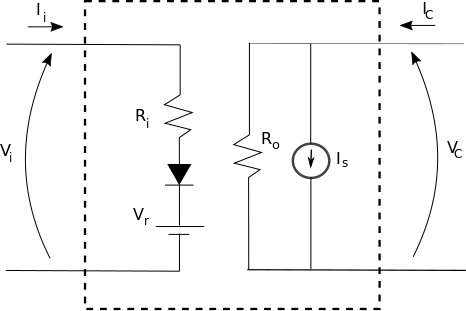}
\caption{The simplified transistor model respective to the NPN transistor characteristic curve in Figure~\ref{fig:parallels} and the
ideal diode model adopted for representing the reversely biased base-collector junction.}
\label{fig:equiv_Rofixed}}
\end{figure}

The simplified characteristic surface in Figure~\ref{fig:parallels}, together with the assumed simplified diode model, yields the equivalent 
circuit shown in Figure~\ref{fig:equiv_Rofixed}.  The ideal diode is included for the generality's sake, but can be omitted under the
adopted transistor base-emitter forward bias.  Observe that the $I_B$-indexed isolines are obtained by incorporating the constant 
current source $I_s$ in parallel with an internal output resistance $R_o$ (Norton equivalent).

\begin{figure}[h!]
\centering{
\includegraphics[width=8cm]{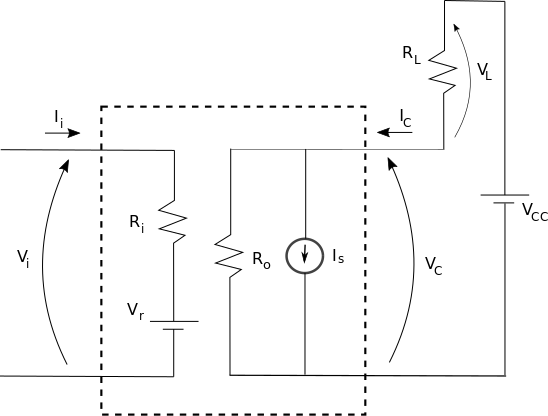}
\caption{The circuit configuration considered in this work.  The load resistance $R_L$, together with a series voltage supply $V_{CC}$, are
attached to the transistor collector (here playing the role of output).  The inclined parallel isolines in the adopted characteristic surface
are modeled in terms of the variable current source $I_s$ and the respective output resistance $R_o$.}
\label{fig:eq_load_fixed}}
\end{figure}

We consider a simplified common emitter circuit configuration as in Figure~\ref{fig:eq_load_fixed}.  Here, a load resistance 
$R_L$ is attached between the external voltage source
$V_{CC}$ and the transistor output (collector).  It can be easily verified that this pair of components external to the transistor 
define a straight line in the $(V_C,I_C)$ space as in Figure~\ref{fig:parallels}.  This line is commonly called \emph{load line}. 
The circuit operation is now restricted to take values
$(V_C,I_C)$ along this load line.  While the resistance $R_L$ corresponds to the actual load of the transistor, the voltage source $V_{CC}$
is required because the transistor, being a passive device, can not deliver power into the load, but only control an external voltage or
current source according to the input current $I_B$.   That is why transistors operating as amplifiers are sometimes called ``\emph{valves}''.

First, we approach the base-emitter junction, which we assumed to operate as a simplified diode.  For this input port, we have:

\begin{equation}
   V_B = V_r + R_i I_B \label{eq:base}
\end{equation}

Now, we proceed to the output port.  By defining the transistor current gain $\beta = I_R/I_s = constant$ and applying Kirchhoff's 
and Ohm's laws (e.g.~\cite{hyat:1962}), we have that:

\begin{eqnarray}
   I_C \left( I_B \right)  = \frac{V_{CC} + \beta R_o I_B}{R_o + R_L}  \\
   V_L \left( I_B \right) =  R_L I_C  =  R_L \frac{V_{CC} + \beta R_o I_B}{R_o + R_L}  \\
   P_L \left( I_B \right)  = R_L  \left( \frac{V_{CC} + \beta R_o I_B}{R_o + R_L}  \right) ^2  
\end{eqnarray}

When $R_0 \rightarrow \infty$, the transistor output stage becomes a perfect current source and we have:

\begin{eqnarray}
   I_C \left( I_B \right)  =  \beta I_B  \\
   V_L \left( I_B \right) =  R_L I_C  =  R_L \beta I_B  \\
   P_L \left( I_B \right)  = R_L  (\beta I_B ) ^2  
\end{eqnarray}

The AC current, voltage and power gains at the resistance load are commonly defined, respectively, as:

\begin{eqnarray}
   a_i (I_B)  = \frac{I_C(I_B)}{I_B}  \bigg|_{Q}\\
   a_v (I_B)  = \frac{V_L(I_B)}{V_i(I_B)} \bigg|_{Q}  \\
   a_p (I_B)  = \frac{P_L(I_B)}{P_i(I_B)} \bigg|_{Q}  =  a_i(I_B) a_v(I_B)  
\end{eqnarray}

Strictly speaking, the AC gains are defined respectively to an operating point $Q = (V_{C,Q},I_{C,Q})$. This effectively
implies $R_i \rightarrow 0$.  In addition, because of the constant slope of the $I_B$-indexed isolines in the simplified transistor 
model under consideration,  the choice of this point becomes immaterial and we have:

\begin{eqnarray}
   a_i  = \frac{\beta R_o }{(R_o + R_L)} \\
   a_v (F)   = \frac{\beta R_L   R_o}{R_i(R_o + R_L)}   \label{eq:av_trad}  \\
   a_p (I_B)  = \frac{R_L  (\beta R_o)^2  }{ R_i (R_o + R_L)^2}   \label{eq:ap_trad}
\end{eqnarray}

So, we have that all the considered gains do not depend on $I_B$, $V_C$ or $I_C$, as would
be expected as a consequence of adopting equispaced isolines. 
For the further simplified case when $R_o \rightarrow \infty$, we have:

\begin{eqnarray}
   a_i  = \beta \\
   a_v (I_B)   = \beta \frac{R_L}{R_i}  \label{eq:av_tradit_roinf}  \\
   a_p (I_B)  =  \beta^2 \frac{R_L }{R_i } \label{eq:ap_tradit_roinf} 
\end{eqnarray}

\section{The Early Model Approach} \label{sec:early}

\begin{figure}[h!]
\centering{
\includegraphics[width=8.5cm]{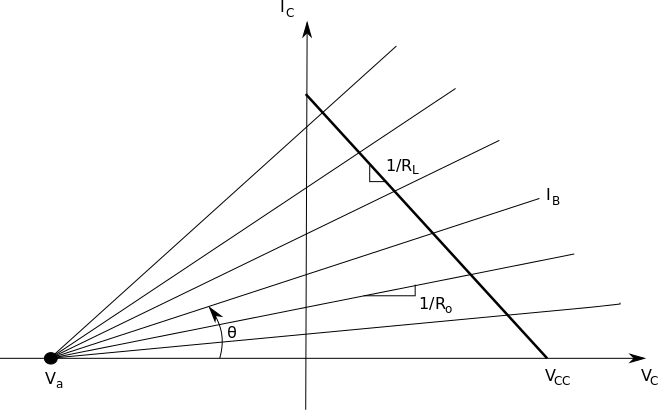}
\caption{The geometrical construction underlying the Early model of transistors.  The set of $I_B$-indexed
isolines converge at $V_C = V_a$ as a consequence of the Early effect.  The load line defined by a resistive
load $R_L$ attached to the collector in series with the voltage supply $V_{CC}$ is also shown.   The circuit operation
therefore remains limited along this line.  Observe that the slope of the load line is determined by $R_L$, while the
inclination of the isolines correspond to $1/R_o(I_B)$, which is intrinsically compatible with the geometric
structure observed for the real-world transistor in Figure~\ref{fig:raw}.  Only a second hypothesis underlies the Early model: the 
linear relationship between the angle $\theta$ of the isolines and the base current $I_B$, i.e.~$\theta = s I_B$,
implying a second parameter, the proportionality constant $s$, to the Early model.}
\label{fig:loadline}}
\end{figure}

Now, we revisit the circuit configuration discussed in the previous section, but adopting a simple Early 
model~\cite{costaearly:2017,costaearly:2018} instead of
the just considered current-source model.   Figure~\ref{fig:loadline} depicts the characteristic surface underlying 
the adopted Early model, defined in terms of a set of $I_B$-indexed isolines radiating from the same point
along the $V_C = V_a$ axis.  Observe the varying  inclination \emph{and} spacing of the $I_B$-indexed characteristic
isolines along the load line, which account for a more accurate and realistic representation of the transistor intrinsic 
electronic properties, as can be immediately inferred by comparing the diagram in Figure~\ref{fig:loadline} with the
experimental characteristic isolines of the real-world transistor in Figure~\ref{fig:raw}.

Though the ``fan''-like radiation may initially appear as a complication, it will turn out that this is not really so.
In addition, this geometrical organization provides a much more accurate representation of the behavior or real-world
transistors than the previous simplified approach, as it \emph{allows the increasing slopes of the non-equally spaced 
isolines to be taken into account}~\cite{costaearly:2017,costaearly:2018}.
Observe that these slopes correspond to the inverse of the output resistance $R_o$, and also that $tan(\theta) = 1/R_o$.  
In addition, the two parameters involved in the Early modeling, namely the Early voltage $V_a$ and the proportionality
parameter $s$, remain \emph{constant throughout the operation space}, i.e.~these two parameters do not depend
on either $V_C$ or $I_C$ (which is not verified for more traditional approaches based on $\beta$ and $R_o$).  
These key properties of the Early approach, that constitute the main motivation of the
present work, are summarized in the following box: \vspace{0.5cm}

\noindent\fbox{
    \parbox{8cm}{
    In the Early model, the characteristic surface of real-world transistors, \emph{allowing for varying slope
    of non-equally spaced characteristic isolines indexed by the base current $I_B$}, is represented in terms 
    of only two \emph{constant parameters} ($V_a$ and $s$) that provide
    a very comprehensive characterization of the device operation.
    }}

\vspace{0.5cm}
The possible reason why such an Early model approach~\cite{costaearly:2017,costaearly:2018} was 
been adopted earlier is because the relationship between $\theta$ and $I_B$ did not seem to
be known.  However, at least for several types of small signal transistors, it has been experimentally verified 
recently~\cite{costaearly:2017,costaearly:2018} 
that $\beta = s I_B$, allowing a substantially simple and effective transistor model to be developed and applied~\cite{costaearly:2017,costaearly:2018}. 
Therefore, only two elements underly the Early model:  the hypotheses that $\theta = s I_B$, and the fact that the all the
$I_B$-indexed isolines intersect at a same value $V_a$ along the $V_C$ axis.  The Early model is immediately applicable
equally to NPN and PNP models, and in this work all PNP currents and voltages are shown with negative values, for
simplicity's sake.

\begin{figure}[h!]
\centering{
\includegraphics[width=8cm]{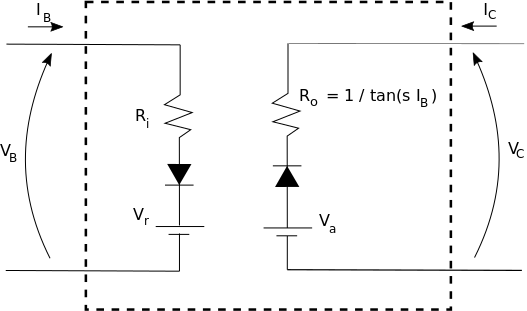}
\caption{The \emph{Early model equivalent circuit}.  While the input port is the same as before, the output port
now incorporates a \emph{variable} output resistance $R_o(I_B) = 1/tan(sI_B)$ and a \emph{fixed} voltage source $V_a$, where $V_a$ is
the Early voltage.  The two diodes are included for generality and can be omitted in typical amplifying circuit configurations.}
\label{fig:equiv}}
\end{figure}

The equivalent circuit of the graphical construction shown in Figure~\ref{fig:loadline} can now be easily derived as illustrated in 
Figure~\ref{fig:equiv}.   The input port is modeled in the same way as in the previous section.
However, in the Early approach, the output port of a transistor is now represented as \emph{voltage} source with \emph{fixed} value
$V_a$ corresponding to the \emph{Early voltage}, in series with a \emph{variable} output resistance $R_o(I_B) = 1/tan(s I_B)$,
where $s$ is the proportionality constant constituting the second parameter in the Early model~\cite{costaearly:2017,costaearly:2018}.  
Rarely, if ever, equivalent
circuits employ variable resistances (or conductances), but there is no intrinsic shortcoming in this type of approach.  On the 
contrary, the $I_B$-controlled resistance of the isolines provides a natural represented by the geometrical organization of the
transistor operation.  Observe also that though a respective Norton version of the equivalent circuit proposed for the Early
modeling would be possible, it would imply both output resistance and current source to vary with $I_B$, while the voltage
source in the adopted Thevenin configuration remains constant with $I_B$ as well as with $I_C$ and $V_C$.

The presented Early equivalent circuit can be understood as having similar complexity to the varying current source model 
in Figure~\ref{fig:equiv_Rofixed}.  However, \emph{it is typically much more accurate than that model as allowed by the variable 
resistance representation of the progressively more inclined, non equally space isolines found in most real-world transistors}.  It could also be
argued that the voltage-based (Thevenin) approach is probably more intuitive than the more traditional  current-based
(Norton) counterpart.  

\begin{figure}[h!]
\centering{
\includegraphics[width=8.5cm]{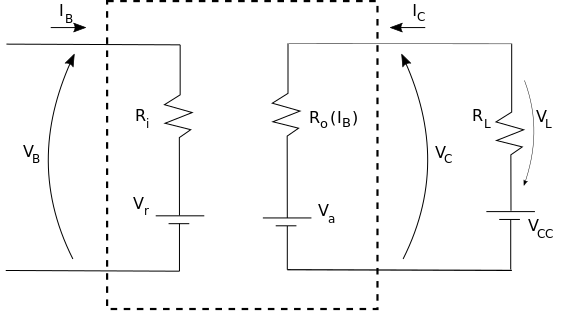}
\caption{Application of the Early model to study an simplified common-emitter configuration with resistive load $R_L$
and voltage source $V_{CC}$.  Recall that $V_a$ has negative value and that the diodes can be omitted for the assumed
biases. While $V_a$ does not vary with either $I_B$, $I_C$ or $V_C$, the proportionality parameter $s$ depends of $I_B$,
implying the output resistance $R_o$ also to be a function of $I_B$.}
\label{fig:earlycircuit}}
\end{figure}

Figure~\ref{fig:earlycircuit} shows the same circuit as in the previous section, but with the transistor represented in terms of its
Early model.  Since the base-emitter is forward biased and the base-collector is inversely biased, the two diodes can be
conveniently omitted.

Because the input port is modeled in the same ways as in the previous section, Equation~\ref{eq:base} is verified also for this
circuit configuration.  The output current, as well as $V_L$ and  $P_L$ at the output stage can be immediately determined by applying 
Kirchhoff and Ohm's laws as:

\begin{eqnarray}
   I_C \left( I_B \right) = \frac{V_{CC} - V_a}{R_L + Ro(I_B)} \\
   V_L \left( I_B \right) =  R_L  \frac{V_{CC} - V_a}{R_L + Ro(I_B)}  \\
   P_L \left( I_B \right)  = R_L \left( \frac{V_{CC} - V_a}{R_L + Ro(I_B)}  \right) ^2  
\end{eqnarray}

where $R_o(I_B) = 1/tan(s I_B)$.  These expressions are hardly more complicated than those obtained for the more traditional, and
less precise, modeling approached addressed in the previous section.  

The current, voltage and power gain are now given as:

\begin{eqnarray}
   A_i (I_B)  =  \frac{(V_{CC} - V_a)tan(s I_B)}{I_B(R_L tan(s I_B) + 1)} \\
   A_v (I_B)   =   \frac{R_L (V_{CC} - V_a)tan(s I_B)}{(R_L tan(s I_B) +1)(V_r + R_i I_B)}     
\end{eqnarray}

Though this last pair of equations is slightly more sophisticated than those obtained for the more traditional modeling approach discussed
in the previous section, this is so because now they account for the important fact that \emph{all gains are indeed functions of $I_B$}.
This is particularly important because this dependence directly implies that the current, voltage and power amplifications
will necessarily be \emph{non-linear} in a real-world circuit employing a real-world transistor.

The AC version of the obtained current and voltage gains (the power gain will be henceforth omitted as it can be directly obtained
as $a_p = a_i a_v$) can now be calculated as:

\begin{eqnarray}
   a_i (I_B)  =  - \frac{V_a tan(s I_B)}{I_B(R_L tan(s I_B) + 1)}  \label{eq:aipre} \\
   a_v (I_B)   =  - \frac{R_L V_a tan(s I_B)}{R_i I_B(R_L tan(s I_B) +1)}   \label{eq:avpre}     
\end{eqnarray}

It is interesting to rewrite the above equations as:

\begin{eqnarray}
   a_i (I_B)  =  - V_a \left[ \frac{ tan(s I_B)}{I_B(R_L tan(s I_B) + 1)}  \right] \label{eq:aipre} \\
   a_v (I_B)   =  - \frac{V_a}{R_i} \left[ \frac{R_L tan(s I_B)}{ I_B(R_L tan(s I_B) +1)}  \right]  \label{eq:avpre}     
\end{eqnarray}

so that we can now define the two functions $g()$ and $h()$ that play the role of \emph{kernels} of the transistor operation:

\begin{eqnarray}
   g(I_B , s,R_L) =   \frac{ tan(s I_B)}{I_B(R_L tan(s I_B) + 1)}   \qquad  \label{eq:g} \\
   h(I_B , s,R_L) =   \frac{R_L tan(s I_B)}{ I_B(R_L tan(s I_B) +1)}   \label{eq:h}  \qquad
\end{eqnarray}

These functions are critically important because they define the linearity of the transistor amplification in the
considered circuit configuration.   Both $g()$ and $h()$ depend only of the base current $I_B$, the proportionality 
parameter $s$, and the load resistance $R_L$.   The current and voltage gains can be easily obtained by 
multiplying $g()$ and $h()$ by $-V_a$ and $-V_a/R_i$, respectively, as implied by Equations~\ref{eq:ai} and~\ref{eq:av}. 

\begin{figure}[h!]
\centering{
\includegraphics[width=8.5cm]{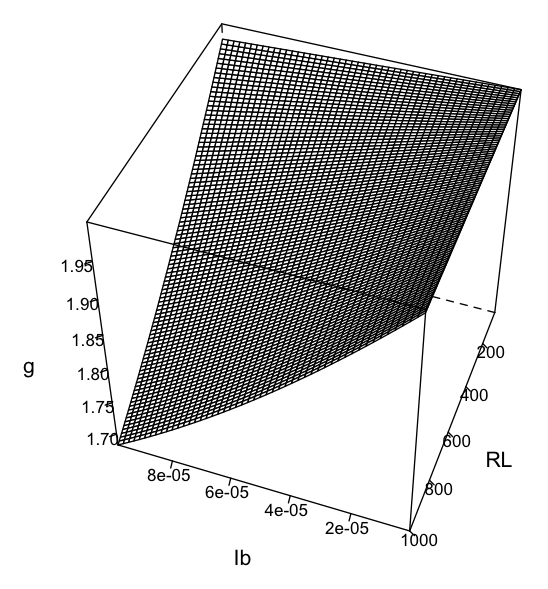}
\caption{The function $g(I_B | s, R_L)$ in terms of $0 \leq I_B \leq 100\mu A$ characterizing the current 
amplification (except by a factor $-Va$)  properties of a transistor for  $s=2$ and $50 \Omega \leq R_L  \leq1 k \Omega$.  }
\label{fig:g}}
\end{figure}

\begin{figure}[h!]
\centering{
\includegraphics[width=8.27cm]{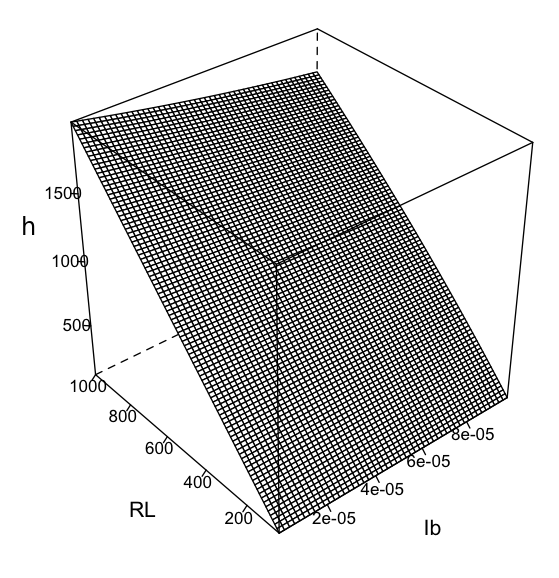}
\caption{The function $h(I_B | s, R_L)$ in terms of $0 \leq I_B \leq 100\mu A$ characterizing the voltage 
amplification (except by a factor $-V_a/R_i$) properties of a transistor for $s=2$ and $50 \Omega \leq R_L  \leq1 k \Omega$.  }
\label{fig:h}}
\end{figure}

Figure~\ref{fig:g} illustrates the behavior of $g()$ in terms of $I_B$, and Figure~\ref{fig:h} depicts the function $h()$
in terms of these same two variables for $0 \leq I_B \leq100 \mu A$ and $R_L=670 \Omega$. For simplicity's sake, we 
henceforth set $V_a=-100$ and $s=2$, which is typical of small signal real-world NPN devices~\cite{costaearly:2018}.  
Both cases assume $s=2$ and $50 \Omega \leq R_L \leq 1 k \Omega$, but similar behaviors are observed for other 
typical parameter value configurations.

For any fixed $R_L$, $g()$ decreases with $I_B$ in an almost linear fashion, especially for smaller values of $R_L$.  
Similarly,  $h()$ also decreases with $I_B$ for fixed $R_L$ (except for $R_L=0$, when $g()=0$), but now
in an almost perfectly linear fashion.  
Indeed, the relative variations of $g()$ and $h()$ with $I_B$ both increase with the fixed $R_L$ value.

Given the almost linear variations of $g()$ and $h()$, they can be effectively characterized in terms of their
average and variation values.  The former of these are immediately given as:

\begin{eqnarray}
   \langle a_i  \rangle \bigg|_{I_{B,min}}^{I_{B,max}} =  \frac{g (I_{B,min}) + g(I_{B,max} )}  {2}  \qquad  \\
   \langle a_v \rangle  \bigg|_{I_{B,min}}^{I_{B,max}} =   \frac{h (I_{B,min}) + h(I_{B,max} )}  {2} \qquad
\end{eqnarray}

The gain variations are all important for characterizing the linearity of the transistor operation (observe that
increased variations will imply larger distortions in the amplification), so that it is worth defining them
in terms of the relative maximum excursions, i.e.:   

\begin{eqnarray}
   \delta_{a_i}  \bigg|_{I_{B,min}}^{I_{B,max}} =  \frac{g (I_{B,min}) - g(I_{B,max} )}  {I_{B,max} - I_{B,min}}  \qquad  \\
   \delta_{a_v}  \bigg|_{I_{B,min}}^{I_{B,max}} =   \frac{h (I_{B,min}) - h(I_{B,max} )}  {I_{B,max} - I_{B,min}} \qquad
\end{eqnarray}

where $g()$ and $h()$ are immediately calculated by using Equations~\ref{eq:g} and ~\ref{eq:h}, respectively.
Interestingly, these variations are directly related to the total distortion implied by the respective amplifications.  
In particular, full linear operation would be achieved if the variations were null.

The nearly linear variations of the current and voltage gain with $I_B$ indicate that, at least for the considered parameters
and variable ranges, the properties of those gains can be effectively quantified in terms of the respective average and variation
values.  

\begin{figure}[h!]
\centering{
\includegraphics[width=9cm]{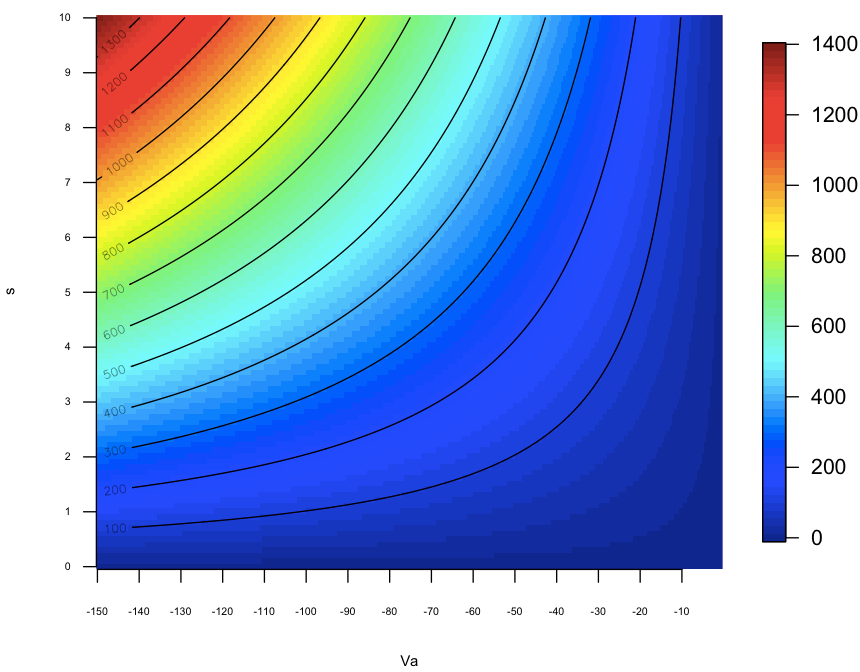}
\caption{The  current gain averaged for $0 \leq I_B \leq 100 \mu A$ for the considered circuit and parameter configurations. 
Respective isolines are also shown.  Though this result was obtained with respect to $R_L670 \Omega$, $V_a=-100$ and $s=2$,
similar structures are obtained for other typical configurations.  Observe that the maximum average gain is achieved at the
upper left-hand side of the Early space, decreasing steadily towards both the $V_a$ and $s$ axes. }
\label{fig:perf_ai_av}}
\end{figure}

We are now in a position to study the overall AC properties of the considered transistor/circuit configuration  --
namely the current and voltage gains, as well as the overall implied distortions -- in terms of all possible
choices of the involved parameters, i.e.~$V_a$, $s$, $R_i$ and $R_L$.   We limit our discussion to the current
gain because an analogue behavior is observed for the voltage gain. Figure~\ref{fig:perf_ai_av} illustrates the average 
current gain, obtained by multiplying $g()$ by $-V_a$.  

The average current gains tend to reach a peak at the top lefthand side of the Early space $(V_a,s)$ and the lowest values
near the $V_a$ and $s$ axes.    Respective isolines, indexed by the average current gain, are also depicted in this
figure.  They present progressively varying shapes, with curvature increasing towards the coordinate system origin.
Observe that the average current gain increases in a non-linear fashion throughout the Early space, being characterized
by steeper variations near the top lefthand side of the represented space.

\begin{figure}[h!]
\centering{
\includegraphics[width=9cm]{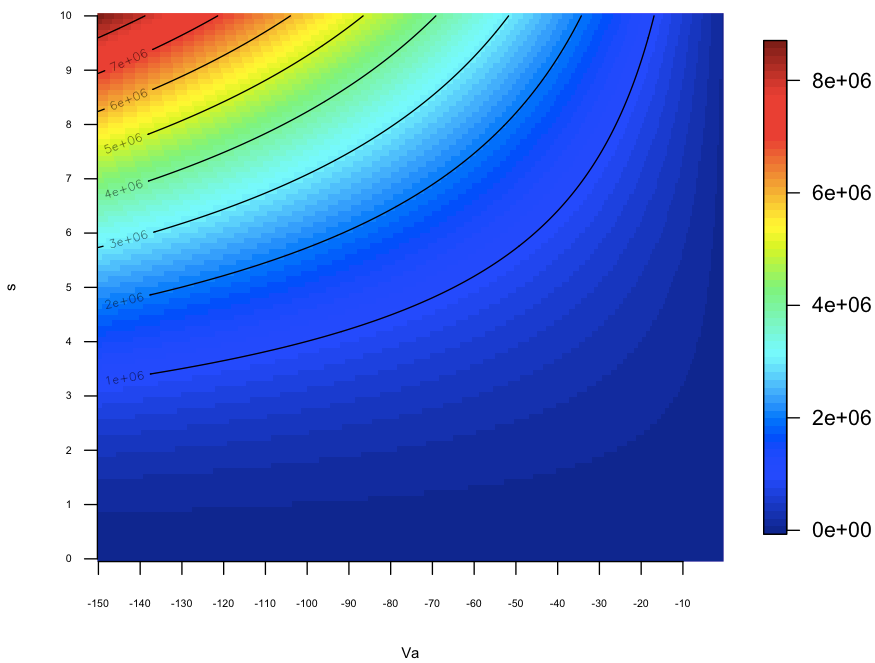}
\caption{The variation of current gain in the interval $0 \leq I_B \leq 100 \mu A$ for the considered circuit and parameter configurations. 
Respective isolines are also shown.  This surface presents similar structure as that obtained for the average current gain in
Figure~\ref{fig:perf_ai_av} }
\label{fig:perf_ai_vr}}
\end{figure}

Figure~\ref{fig:perf_ai_vr} shows the absolute variation of the current gain along the considered portion of the Early
space for $0 \leq I_B \leq100 \mu A$ and $R_L=670 \Omega$.  The obtained pattern is similar to that exhibited by the
average current gain, but it is less symmetrical along the horizontal and vertical orientations.  The respective current
gain variation isolines are also shown in this figure.

\begin{figure}[h!]
\centering{
\includegraphics[width=9cm]{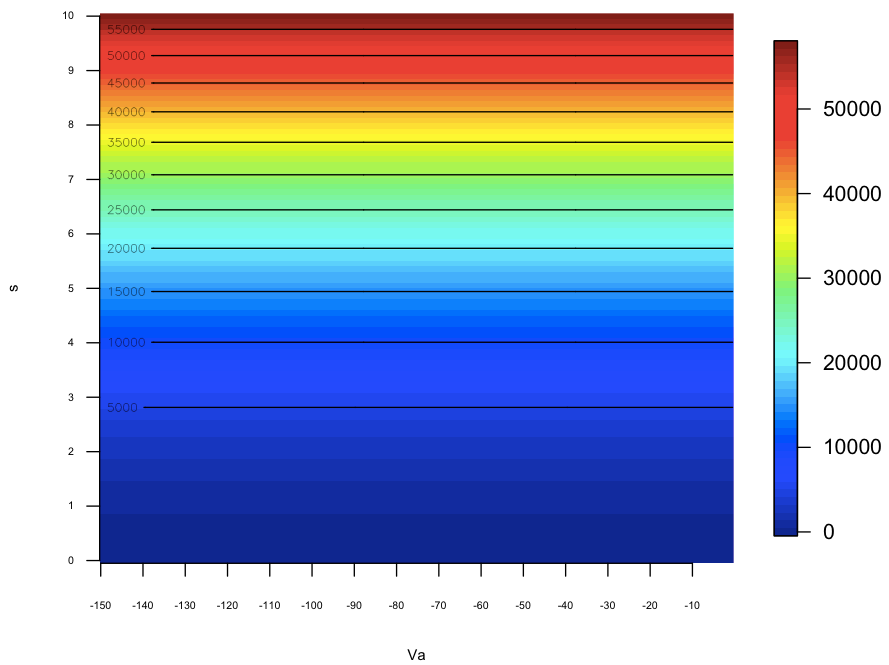}
\caption{The function $h(I_B | s, R_L)$ in terms of $0 \leq I_B \leq 100\mu A$ characterizing the voltage 
amplification (except by a factor $-V_a/R_i$) properties of a transistor for $s=2$ and $50 \Omega \leq R_L  \leq1 k \Omega$.  }
\label{fig:perf_ai_vr_norm}}
\end{figure}

Enhanced linearity is achieved for the $(V_a,s)$ configurations near the $V_a$ or $s$ axes.  Observe that the variation values
are high because the variation is normalized by the $I_B$ interval, which is equal to $100 \mu A$.   Another way to
quantify non-linearity as related to the current gain variations is by considering relative instead of absolute values.   This
type of quantification is adopted henceforth in this work.
Figure~\ref{fig:perf_ai_vr_norm} depicts the relative current gain variations, obtained by dividing the previous variations by the
respective average current gain.  

This remarkable result, already hinted by previous total harmonic distortion considerations in a previous work~\cite{costaearly:2018}, indicates
that \emph{the linearity of the transistor in the adopted circuit depends only of the $s$ values}, being completely independent
of the $V_a$ values.  Therefore, horizontal isolines are obtained in the figure.  Observe that the isolines spacing
decreases progressively with $s$.   The most non-linear amplification is obtained for large values of $s$, irrespectively
of $V_a$.  Interestingly, a (welcomed) substantially wide band of low variation is observed near the $V_a$ axis.

\begin{figure}[h!]
\centering{
\includegraphics[width=8.5cm]{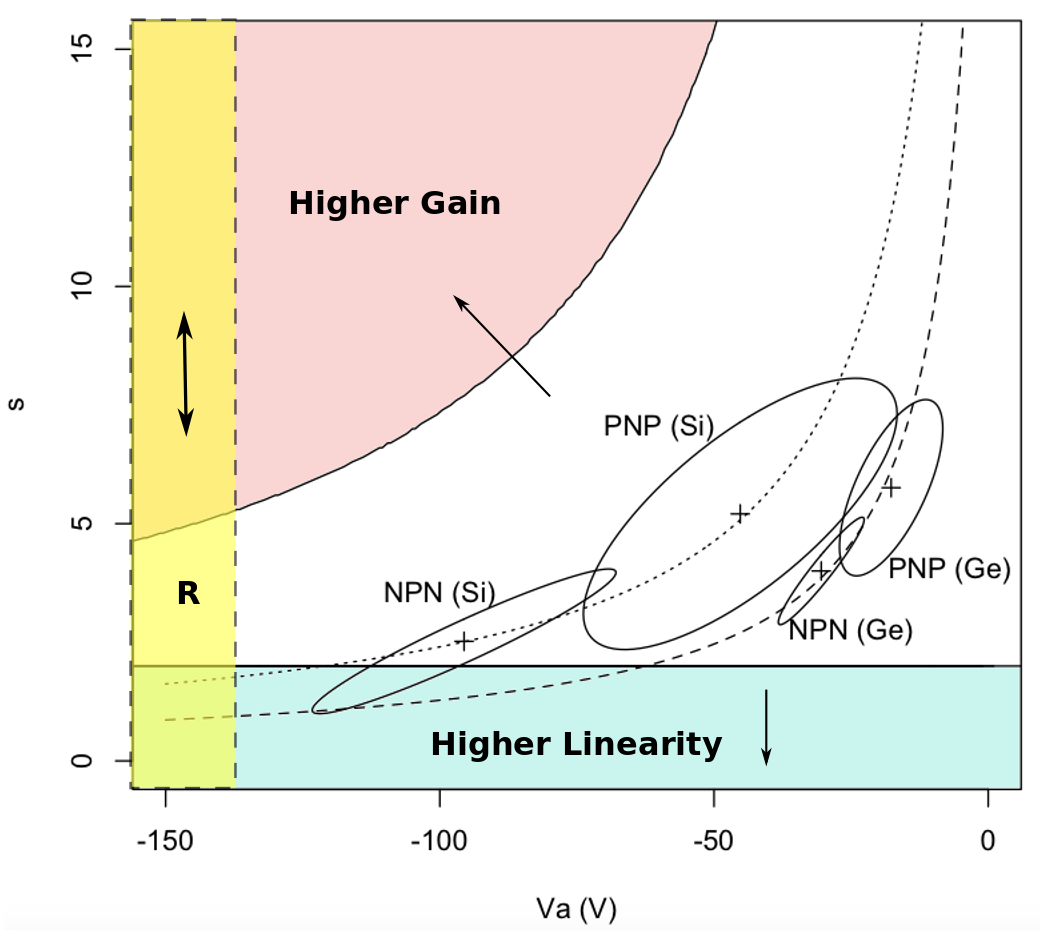}
\caption{The Early space prototypic space, illustrating the high gain and high linearity regions, as well as prototype groups
of NPN and PNP silicon (Si) and germanium (Ge) junction transistors.  Though a trade-off results regarding the choice of
high gain and high linearity, the region $R$ provides a good compromise between these two goals.  Also, the fact that the
higher gain isolines approach the $V_a$ axis for larger values of $V_A$, therefore leading to respectively smaller values
of $s$, also implies that transistors with large values of
$V_a$ provide a good combination of gain and linearity.}
\label{fig:perform}}
\end{figure}

The results revealed by the average current gain and its variations can now be neatly combined into a same diagram, henceforth
called the prototypic Early space, to provide
an overall representation of the trade-off between current gain and linearity in the considered circuit.  Figure~\ref{fig:perform}
shows the so obtained prototypic space, as well as Mahalanobis ellipses obtained experimentally for several NPN and 
NPN silicon~\cite{costaearly:2018} and germanium~\cite{germanium:2018} junction transistors.  The isolines for 
$\beta=130$ (dashed line) as well as for $\beta=240$ (dotted line)
are also shown in the figure, passing very near to the centers of mass (averages) of the considered groups of transistors.

This prototypic space summarizes several key information about transistor operation, providing valuable subsidies for
choice, applications, and design of transistors.  If priority is placed on gain, devices near the red region should be chosen.
Contrariwise, in case linearity is the main objective, devices near the green band should be considered.  Now, it becomes
evident that the high gain and high linearity tend to follow contrary pathways along the prototypic space, implying a trade-off between
these two often sought characteristics.  However, the different geometry of the average current gain and variation surfaces 
imply some regions of special interest when trying to optimize both gain and linearity.  Though, in principle, the linearity
does not depend of $V_a$, if a large magnitude value of $V_a$ is chosen, such as in the yellow band $R$ in the figure, higher 
gain values can be obtained even when choosing transistors with smaller values of $s$.  In this way, the yellow region stands as
having particular interest when aiming simultaneously at hight gain and good linearity.  Also, observe the higher gain
isolines tend to approach the $V_a$ axis for large magnitude values of that parameter, the implied funneling effect also implies 
lower values of $s$ and improved linearity.  Thus, devices with very large $V_a$ are particularly interesting for providing a
good combination of high gain and low distortion.

Interestingly, real-world NPN and PNP transistors, at least for the cases experimentally characterized~\cite{costaearly:2018}, tend to
occupy an intermediate position in the Early parameter space.  The silicon NPN devices are those nearer to the
yellow band, representing good candidates when aiming at both high gain and low distortion levels.

\section{Stability Factor Analysis}

An important property of transistor-based amplifying circuits regards their \emph{stability} (e.g.~\cite{tinnell:1972}) with
respect to the current or voltage supply oscillations.  Let's obtain, by using the  Early equivalent circuit,
the current and voltage stability for $V_{CC}$ oscillations of the circuit considered in the previous sections.   
This can be done as follows with respect to the collector current and voltage instability with respect to 
$V_{CC}$ oscillations:

\begin{eqnarray}
   S_i = \frac{\partial{I_C}}{\partial V_{CC}} \left( I_B \right)  = \frac{1+ \beta R_o I_B}{R_o + R_L}  \\
   S_v = \frac{\partial{V_L}}{\partial V_{CC}} \left( I_B \right) =   R_L \frac{1+\beta R_o I_B}{R_o + R_L}    
\end{eqnarray}

It follows that:

\begin{eqnarray}
   S_i = \frac{\partial{I_C}}{\partial V_{CC}} \left( I_B \right)  =      \frac{(1-V_a)tan(sI_B)}{R_L tan(sI_B)+ 1}   \\
   S_v = \frac{\partial{V_L}}{\partial V_{CC}} \left( I_B \right) = R_L \frac{(1-V_a)tan(sI_B)}{R_L tan(sI_B)+ 1} 
\end{eqnarray}

\begin{figure}[h!]
\centering{
\includegraphics[width=8.27cm]{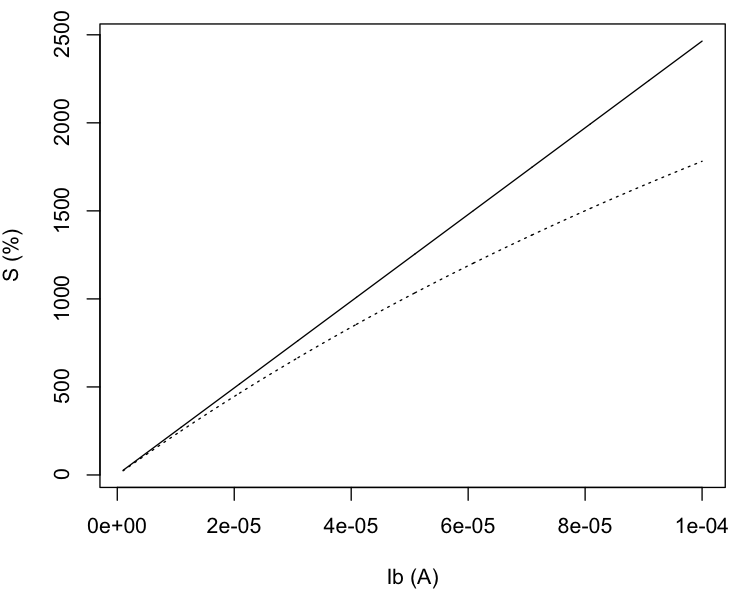}
\caption{The dashed line shows the instability $S$ of voltage in terms of $I_B$ for the considered common emitter 
circuit configuration assuming $V_a = -70V$ and $R_L = 670 \Omega$.  The  solid line corresponds to the
voltage stability as calculated by using the more traditional alternative model.  The two curves deviated markedly,
specially for larger values of $I_B$.}
\label{fig:S}}
\end{figure}

The voltage instability $S_v$ is directly related to the current stability $S_v$, so we focus
on the former index.  Figure~\ref{fig:S} illustrates the behavior of $S_i$ in terms of $0 \leq I_B \leq 100 \mu A$
assuming $V_a = -70V$ and $R_L = 670 \Omega$.  The $\beta$ and $R_o$ used in the more traditional approach
in Section~\ref{sec:trad} were obtained by using the mapping between the Early and $(\beta, R_o)$ parameters developed 
in~\cite{costaearly:2018}.  

A substantial discrepancy can be observed between the instability curves obtained by using the Early (dashed lines) and
more traditional approach (solid line), as these diverge markedly, specially for larger values of $I_B$.  Interestingly,
the instability obtained by using the more accurate Early equivalent circuit is smaller than that yielded by using the
more traditional model.  This is a probable consequence of the fact that the varying slopes of the $I_B$-indexed isolines,
which is represented in the Early equivalent circuit but not the more traditional approach, compensate for the
voltage supply oscillations through a kind of saturating effect.  Thus, real-world devices may turn out to have potentially
increased stability than that revealed by more traditional approaches based on the $(\beta,R_o)$ parameters.

This simple application of the Early equivalent circuit to the analysis of an important practical parameter in linear
circuit design and application, namely the variation of the amplification as a consequence of power supply 
oscillations, revealed that substantial discrepancies in the quantified electronic behavior of the analyzed circuits
can be produce when overlooking the more realistic variation of the $I_B$-indexed isolines in real-world devices.

\section{Parallel Transistor Configurations}

\begin{figure}[h!]
\centering{
\includegraphics[width=8cm]{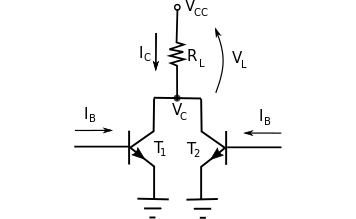}
\caption{Parallel combination of two transistors that is sometimes employed to deliver more current 
to the load and to reduce the output resistance.  This is not a practical circuit as it lacks some
important considerations.  }
\label{fig:eqpar}}
\end{figure}

A single transistor is capable of delivering a limited amount of current to the load, being subjected to maximum
absolute ratings.  A possible means for trying to increase the power provided by the output port of  transistors
is to combine them in parallel as illustrated in Figure~\ref{fig:eqpar}.

\begin{figure}[h!]
\centering{
\includegraphics[width=8.5cm]{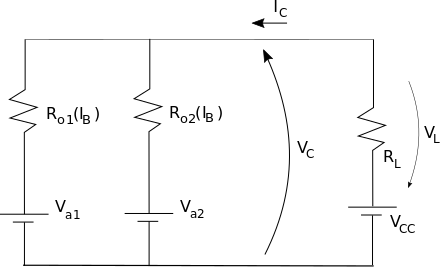}
\caption{The circuit obtained by substituting the two transistor in the parallel combination by their
respective Early equivalents. }
\label{fig:ptrans}}
\end{figure}

As illustrated in Figure~\ref{fig:ptrans}, the Early equivalent circuit can be easily employed to analyze 
the parallel configuration of two transistors (the result can be immediately extended to more transistors).  

The so obtained circuit can now be analyzed by using Kirchhoff's and Ohm's laws, as well as the
Thevenin equivalent theorem, to obtain the following equations of the resulting effective output
resistance $R_{o,eq}(I_B)$ and Early voltage $V_a(I_B)$, both of which potentially dependent
of $I_B$:

\begin{eqnarray}
   R_{o,eq}(I_B) = \frac{R_{o1}(I_B) R_{o2}(I_B)}{R_{o1}(I_B) + R_{o2}(I_B)}  \\
   V_{a,eq}(I_B) = V_{a2} + R_{o2} (I_B) \frac{V_{a1} - V_{a2}}{R_{o1}(I_B) + R_{o2}(I_B)}  
\end{eqnarray}

For simplicity's sake, and without loss of generality, we can make $s_{2} = \gamma s_{1}$ and $V_{a2} = \lambda V_{a1}$.  
First, we have that:

\begin{equation}
  \frac{R_{o2}(I_B)}{R_{o1}(I_B)} = \frac{ tan(s_{1} I_B)}{tan(\gamma s_{1} I_B)}
\end{equation}

and, consequently, it follows that:

\begin{eqnarray}
   R_{o,eq}(I_B) = \frac{1}{tan(s_1 I_B) + tan(\gamma s_1 I_B)}  \label{eq:Roeq} \\
   V_{a,eq}(I_B) =  \frac{\lambda  V_{a1} tan(s_1 I_B) + V_a tan(s_1I_B)}{tan(s_1 I_B) + tan(\gamma s_1 I_B)}  \label{eq:Vaeq}
\end{eqnarray}

Interestingly, $V_{a,eq}$ turns out to be \emph{almost perfectly constant with} $I_B$ for the parameters and variables
typically found in the considered circuit configurations to the point that the variation often results smaller than
double floating point precision.  So, the equivalent transistor effectively has, effectively, a \emph{constant Early voltage}
determined only by $s_1$, $s_2$, $V_{a1}$ and $V_{a2}$. This nearly constant value of $V_{a,eq}$ can be conveniently 
obtained by applying L'Hospital theorem to Equation~\ref{eq:Vaeq}, which yields:

\begin{equation}
   V_{a,eq}(I_B) = \frac{\lambda+1}{\gamma +1} V_a     \label{eq:RoeqLho} 
 \end{equation}
 
\begin{figure}[h!]
\centering{
\includegraphics[width=7cm]{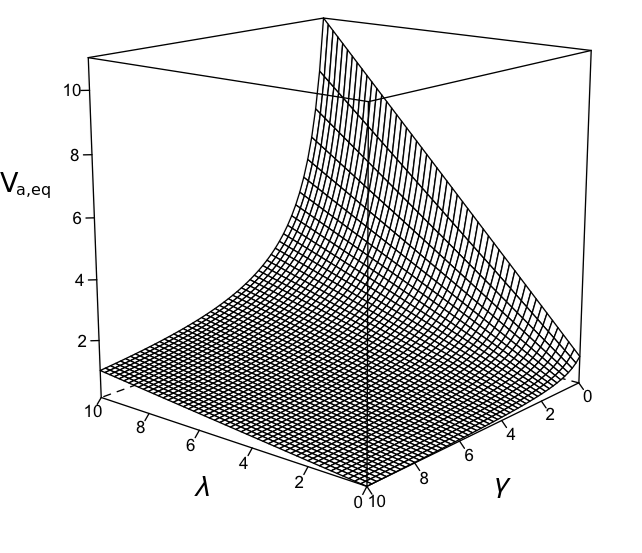}
\caption{The equivalent Early voltage $V_a$ in terms of $\gamma = s_2/s_1$ and $\lambda = V_{a2}/V_{a1}$. 
$V_{a,eq}$ increases linearly with $\lambda$ for any fixed value of $\gamma$. It is assumed that $R_{o1} = 1$
and that $V_{a1} = 1$.}
\label{fig:vaeq}}
\end{figure}

 The dependence of $V_{a,eq}$ with $\gamma$ and $\lambda$ is depicted in Figure~\ref{fig:vaeq}.

So, increasing $\lambda$ promotes linear larger values of $V_{a,eq}$, and this increase is more accentuated for
smaller values of $\gamma$.  However, the effective $V_{a,eq}$ can never exceed the maximum between the
two original Early voltages $V_{a1}$ and $V_{a2}$.

Another remarkable property of the parallel transistor combination is that the proportionality ratio $R_{o1}/R_{o,eq}$ 
and $R_{o2}/R_{o,eq}$ is also virtually invariant with $I_B$.  This means that the pairwise combination of
transistors will have an equivalent output resistance $R_{o,eq}$ that will not interfere with the degree of linearity of the 
resulting device.  As a consequence, parallel combinations of transistors cannot imply in new types of
characteristic surfaces deviating from the Early model geometry.  Similarly, by applying L'Hospital theorem:

\begin{equation}
   R_{o,eq} = \frac{\gamma}{\gamma+ 1} R_{o2}    \label{eq:RoeqLho} 
 \end{equation}

\begin{figure}[h!]
\centering{
\includegraphics[width=7cm]{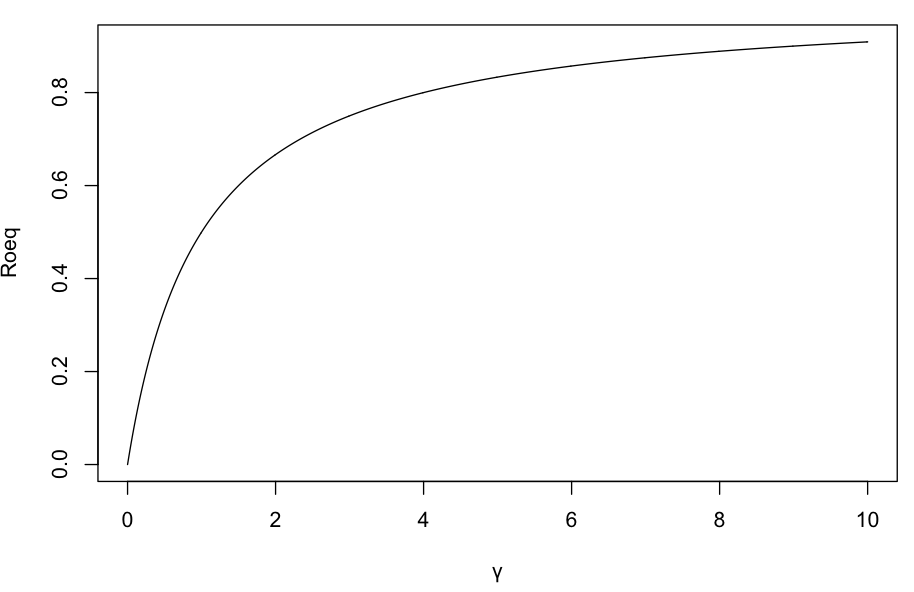}
\caption{The equivalent output resistance $R_{o,eq}$ in terms of $\gamma = s_2/s_1$. 
$R_{o,eq}$ increases with $\gamma$, but tends to saturate soon.  It is assumed that $R_{o1} = 1$.}
\label{fig:gammareq}}
\end{figure}

Figure~\ref{fig:gammareq} shows the range of $R_{o,eq}$ that can be obtained by varying $\gamma$.  Observe that
the equivalent resistance tends to saturate at 1 for very large values of $\gamma$.

\begin{figure}[h!]
\centering{
\includegraphics[width=8cm]{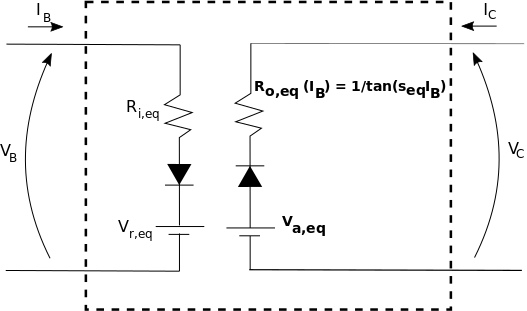}
\caption{The resulting Early equivalent circuit for parallel combination of two transistors.  The resulting 
device has been found, at least for the considered configurations, to follow almost perfectly the Early
model in the sense that the resulting equivalent Early voltage is constant and the obtained equivalent
output resistance depends on $I_B$ as dictated by the Early approach.  }
\label{fig:pareq}}
\end{figure}

As the equivalent $V_a$ and $s$ initially seemed to depend on $I_B$, this could lead to novel qualitative behavior
of the parallel combination of transistors, such as eventual displacement of the point $V_a$ along the $V_C$ axis
during normal circuit operation, or non-linear dependencies of $\theta$ with $I_B$ implied by $s$ not being constant with $I_B$.
However, the interesting obtained results imply that the parallel combination of transistors, at least for the considered 
configuration and parameters and variables ranges, will effectively result in a new transistor described by the same qualitative
behavior as dictated by the Early equivalent circuit (see Figure~\ref{fig:pareq}).  So, parallel combinations can be 
used to ``design'' new devices
with parameters that are intermediate between the two original transistors.  For instance, a transistor with smaller $V_{a,eq}$ can
be obtained by combining two transistors with larger $V_a$ values.  Observe that the resulting $V_{a,eq}$ and $R_{o,eq}$
values will never be larger than the respective values of the original combined transistors.  

\section{Concluding Remarks}

Transistors have been around since the mid 40's, playing a decisive role for the development and popularization of modern
electronics.  Yet, their seemingly complex operation, which constrains the linearity of applications, has motivated much effort aimed
at better understanding, modeling, applying and designing improved devices and circuits.   There are two main vectors motivating
the continued interest in transistors: (i) they are used in a vast range of applications underlying most of human activities, some
of them critically; and (ii) oftentimes, transistors are required to operate with as much linearity as possible, which implies several
challenges given the practical variability of transistor parameters allied to their relatively complex behavior.    Originally, the focus
of interest in transistor research was directed toward discrete devices and circuits, and great attention tended to be given to
graphical representation and modeling approaches.  With the introduction of integrated technology,
interest shifted to transistors as the basic elements in integrated circuits, often studied by using numeric-computational simulations.  
However, several applications still justify, or even require
(as in power electronics), the application of discrete devices.  Be that as it may, discrete and integrated transistors follow the
same physical constrains, so that they can be treated by the same unified approaches.

As a consequence of the continuing interest in transistor electronics, a vast quantity of works have been reported aimed at
characterizing, modeling and design these devices.  One of the main approaches consists in deriving, experimentally or from
basic physical principles, the functional response of the devices (often done in a graphic way, with the help of transfer functions and
other types of diagram), and then obtaining respective equivalent circuits that can be incorporated into circuit analysis performed
analytically or through simulations.  Several of the traditional transistor models are based on two parameters, the current gain
$\beta$ and the (collector) output resistance $R_o$.  While these two parameters are particularly intuitive, as they are closely
related to transistor electronic operation in circuits, they have an intrinsic limitation in the sense that they vary, often significantly,
with normal circuit operation, taking different values for each current and voltage values taken by the transistor output.

A simple, and yet accurate, transistor model was developed recently~\cite{costaearly:2017,costaearly:2018} founded on the 
Early effect, characterized by an interesting geometry of operation.  In addition to the
Early voltage $V_a$ sometimes used in transistor modeling, a second proportionality parameter $s$ is involved that governs
how the angle $\theta$ of the base current-indexed transistor isolines vary with that current $I_B$.  Interestingly, this parameter $s$ 
has been experimentally found~\cite{costaearly:2017,costaearly:2018,germanium:2018} to underly the simple linear 
relationship $\theta = s I_B$, at least for hundreds of small signal 
silicon and germanium BJTs, to vary in nearly full linear fashion.  Therefore, both parameters $V_a$ and $s$ are fixed and
constant for each given transistor, defining to a great extent its electronic properties.  In addition, the mapping of transistors into
the Early parametric space, instead of in more traditional $(\beta,R_o)$ spaces, has the advantage of allowing a better 
characterization and separation of different types of transistors~\cite{costaearly:2018,germanium:2018}, possibly as a 
consequence of the fact that the Early approach captures directly the geometry of operation of real-world devices in the 
so-called ``linear'' operation region.

Despite having been introduced quite recently, the Early model has already been applied with encouraging success to several
problems in electronics, including the estimation of transfer functions~\cite{costaearly:2017}, the characterization of complementary BJTs including
the derivation of a prototypic space~\cite{costaearly:2018}, as well as the study of the electronic properties of alloy junction germanium
transistors~\cite{germanium:2018}.

In this work, an equivalent circuit is developed for the Early model in order to pave the way to its direct application in analysis and design of
discrete and integrated circuits.  We started by discussing the geometrical characteristics of real-world transistors, and then
revised one of the simplified traditional models based on $\beta$ and $R_o$, that have been largely used and applied in electronics.
Next, the equivalent circuit of the Early model was derived, starting from characteristic surfaces with isolines radiating from a common
point $V_a$, corresponding to the Early voltage.  The obtained circuit resulted substantially
simple, with its output port including only a resistance and a voltage source connected in Thevenin series.  However, this resistance
turns out to be \emph{dependent on the base current} $I_B$, as well as on the Early parameters $V_a$ and $s$, i.e. $R_o = R_o(I_B | V_a,s)$.

To illustrate the potential of this model, it was applied to the characterization of a simplified common emitter circuit configuration in
which the load is directly attached between the transistor collector and the external voltage supply $V_{CC}$.  The collector current,
as well as the voltage across the load, were easy and conveniently derived by taking into account the Early equivalent circuit.
Interestingly, these equations turned out to be no more complex than those obtained for the more traditional approach based on 
$\beta$ and $R_o$, despite the fact that that the more traditional model considers a much more simplified situation in which the 
current gain does not vary during the transistor operation as a consequence of the equispaced parallel (though inclined) 
isolines. Then, equations for the DC and AC current, voltage and power gains were derived and, again, resulted not to be more elaborated 
than the respective counterparts obtained for the simpler, more traditional model. 

The current and voltage gain equations (the power gain can be directly obtained from these to gains) were found to exhibit
an interesting mathematical structure involving ratios between tangents.  Two kernel functions $g()$ and $h()$ were derived, respectively, 
from these two gain equations.  It is believed that these two equations are behind the intricacies of transistor operation.
As illustrated graphically, the current and voltage gains vary in an almost perfectly linear fashion with $I_B$, which suggests the
derivation of the average and variation of the gains by taking into account only the extremities of the excursion of the gains
as they vary with the same range of $I_B$.  This yielded two very simple expressions for quantifying the average current
and voltage gains, as well as their variations, given the Early parameters $V_a$ and $s$, the transistor input port resistance $R_o$,
as well as the circuit configuration (i.e.~$R_L$).  The gain variations are particularly interesting because they are directly related
to the \emph{linearity} of the amplification.  For instance, null gain variation necessarily implies perfect linearity.  

The development of equations for the average and variation of the current and voltage gains paves the way to a series of interesting
applications and analysis of discrete and integrated circuits.  This potential was preliminary illustrated in the current work with
respect to: (a) the identification of trends, along the Early space, underlying gain and linearity; (b) the derivation of a prototypic
space that can assist transistor characterization, choice, design, and applications; (c) the quantitative study of circuit stability analysis;
and (d) the characterization of parallel configurations of transistors.  

Regarding the gain and linearity of transistors, as revealed by the Early approach, we found that they follow a trade-off, with 
higher gain typically implying reduced linearity.  However, it was possible to identify a region in the Early space in which a good
compromise can be achieved between these two requirements.  In addition, we also found that larger values of $V_a$ magnitude
tend to be particularly interesting, as larger gains can be obtained for a fixed $s$ when $V_a$ is increased while relatively small
values of $s$ can be achieved.  The obtained prototypic
Early space for several real-world NPN and PNP silicon and germanium transistors revealed that the silicon NPN BJTs are the
devices characterizing a particularly good combination of gain and linearity, at least as far as the considered configuration and
parameters and variable settings are concerned.  The obtained prototypic Early space provides valuable means for characterizing, 
analyzing, designing and applying discrete and integrated transistors.  

The application of the Early equivalent circuit to the stability study of effects of power supply oscillations on the load voltage revealed
that the stability of real-world devices in the considered circuit type and configuration may be substantially better than that revealed by
the more traditional modeling approach based on $\beta$ and $R_o$.  This fact suggests that deviations between these two models
can also be expected regarding other important electronic properties of several types of circuits.  Indeed, the Early equivalent
circuit allowed the variation of the slopes $I_B$-indexed isolines to be taken into account, and this has been somewhat provided
some compensation for the power supply oscillations.

Parallel transistor configurations were also approached by using the proposed Early circuit.  Here, we found that the resulting 
device follows almost perfectly the Early representation, as the resulting Early voltage does not depend on $I_B$ and the 
resulting equivalent output resistance depends on that current according to the Early voltage hypothesis, namely
$R_{o,eq} = s_{eq} I_B$.  Such combinations provide an interesting possibility for engineering ``new'' equivalent devices
with intermediate parameters as the two original devices. These results can be immediately extended to parallel combinations
of 3 or more transistors.

The reported concepts, developments, and results pave the way to a large number of immediate future developments.  These
include the investigation of more sophisticated circuit configurations, as voltage followers, current mirrors, push pull stages, phase splitters, 
as well as the more complete common emitter configuration.  Specially promising is the possibility to use the Early equivalent
circuit for developing new, improved circuits. It would also be interesting to compare the performance of several
types of transistors in these circuits by using the Early approach, so as to identify their respective main advantages.  The Early 
equivalent model can also be used to investigate the effect of several types of noise and environment variations on the
circuit performance.  A particularly promising would be to characterize transistor amplification when applied to reactive
loads.  All these possibilities can be used in both discrete and integrated device and circuit design.
\vspace{0.7cm}

\appendix*

\section{Series Expansion of Early Model Current and Voltage Gains}

For typical values of $I_B$ in small signal transistors (approximately in the order of tenths or hundreds of microamperes), the 
Early-based equations obtained for the circuit considered in Section~\ref{sec:early} can be simplified with very small error by respective 
series expansion at $I_B = 0$:

\begin{eqnarray}
   a_i (I_B)  \approx  -  V_a s \left[  1 -  s R_L   I_B + \frac{s^2}{3} (3 R_L^2+1) I_B^2   \right]  \qquad \label{eq:ai}  \\
   a_v (I_B)   \approx   - \frac{ V_a}{R_i} s R_L \left[  1 -  s R_L   I_B + \frac{s^2}{3} (3 R_L^2+1) I_B^2   \right]      \qquad \label{eq:av}
\end{eqnarray}

Often, a good fit can be obtained even by leaving the last term out.  It should be also observed that, when $I_B$ is very
small and $s$ is not very large, we can also make the approximation $s I_B = tan(s I_B)$ in several of the equations in this
present work.

\vspace{0.7cm}
\textbf{Acknowledgments.}

Luciano da F. Costa
thanks CNPq (grant no.~307333/2013-2) for sponsorship. This work has been supported also
by FAPESP grants 11/50761-2 and 2015/22308-2.
\vspace{1cm}


\bibliography{mybib}
\bibliographystyle{plain}
\end{document}